# TECHNOLOGICAL AGGLOMERATION AND THE EMERGENCE OF CLUSTERS AND NETWORKS IN NANOTECHNOLOGY

**Douglas K. R. Robinson[1], Arie Rip[1] and Vincent Mangematin[2]**

1 Department of Science, Technology, Health and Policy Studies, University of Twente, The Netherlands
2 Grenoble Ecole of Management and GAEL (INRA/UPMF), France
Correspondence address: d.k.r.robinson@utwente.nl

Research and development at the nanoscale requires a large degree of integration, from convergence of research disciplines in new fields of enquiry to new linkages between start-ups, regional actors and research facilities. Based on the analysis of two clusters in nanotechnologies (MESA+ (Twente) and other centres in the Netherlands and Minatec in Grenoble in France), the paper discusses the phenomenon of technological agglomeration: co-located scientific and technological fields associated to coordinated technology platforms to some extent actively shaped by institutional entrepreneurs. Such co-location and coordination are probably a pre-requisite for the emergence of strong nano-clusters.

## 1 Introduction

There is a rich literature on high-tech clusters and districts. Case studies have been done, comparisons have been made, and general (even if tentative) conclusions have been formulated, e.g. the role of centres of excellence and star scientists (Zucker et al., 1998, 2002), the size of the existing market (Feldman and Ronzio, 2001, Autant-Bernard et al., 2006) or the role of incumbents and large firms (Agrawal and Cockburn 2003). These studies have often taken biotechnology as their entrance point.

There is an additional dynamic, which we will provisionally call 'technological agglomeration' *i.e.* the geographic co-location of different scientific and technological fields. Technological opportunities as well as requirements on further technological development (e.g. a next generation of chips) stimulate linkages and coordination amongst different fields, and this may create cumulative advantages for clusters in which a wide range of scientific areas is explored. Thus, there is a technological driver in the agglomeration of actors and activities in a geographical region, and more generally, in clusters building on proximity.

Technological agglomeration is a general phenomenon, but it is particularly visible in newly emerging nanotechnology-linked developments. We will use our ongoing studies of regions with a high concentration of nanotechnology-linked activities to show the importance of technological agglomeration for the overall dynamics of development. Our analysis of these techno-institutional dynamics and related changes in networks of firms, research centres, and regional actors and policy makers, takes technology infrastructures and in particular, technology platforms as the main entrance point. Technology platforms are increasingly recognized as important in enabling innovation, as a key part of business models of (high-tech) start-ups, and as having dynamics and requirements of their own.

In this note, we present a first analysis of the role of technological agglomeration in the evolution of nano-clusters in the Netherlands and in Grenoble.

The research note contributes first to the empirical understanding of how technological characteristics are leading to geographic agglomeration of scientific activities. It specifically highlights the role of technological platforms in the agglomeration process. Second, it presents two different processes of agglomeration, a centralised one in France and a distributed one in the Netherlands. Third, our note illustrates the multilevel character of such technological agglomeration.

## 2   The Technological agglomeration and technology platforms

The past ten years have seen an explosion of interest for the area of science and technology labelled "nanotechnology". Nanotechnologies are defined as technologies which include components that have at least one dimension between 1-100 nm, and display unique characteristics due to being at this scale. Unlike previous high-technology waves, nanotechnology covers a diverse field of sciences and engineering, crosses boundaries between them and aims to utilize the very fundamental characteristics of matter by manipulation and control at the nanoscale.

As they cross many disciplines, also many industries and technology chains, nanotechnologies reshape the existing organisational arrangements amongst actors. Technological agglomeration *i.e.* the co-location of scientific and technological supports the development of nanotechnologies within the area. They also involve large investments in infrastructures. Bigger and better clean rooms, atomic force microscopes for observation and manipulation at the nanoscale, e-beam lithography and nano-imprint lithography to make the channels, pores, and circuits needed for the research. Organisationally, it requires the sharing of facilities, equipment and skilled technicians for these very different technology/research fields. Since such facilities are expensive and take some time to construct, they need high investment (both financially and in training of manpower) over a period of time.[1]

---

[1] An example would be the state-of-the art Extreme Ultra-Violet lithography platform which is priced in the order of $40 million (ASML 2005).

- 3 -

Developments in most fields of nanotechnologies are tied to technical facilities, that is the instrumentation itself and the skills that are needed to operate them. In addition, a lot of nanotechnology research involves development, construction and implementation of new instruments. In other words, nanotechnology must be a field that allows us to study the phenomenon of technological agglomeration.

Actually, the infrastructural requirements add up to a basic set of technologies and skills, which allow, when in place, a variety of further work and product development. In other words, there is a technological platform *i.e.* a set of instruments which enables scientific and technological production: it allows exploration and exploitation of a variety of options, for strategic research, technology development, and sometimes also product development. Such a basic set of technical infrastructure is somewhat independent of the team which originally built and assembled it. It is recognized by others as important, and assembled to be able to profit from the variety of purposes it can be put to. It is not focused, however, on appropriating part of the value added in producing goods or services, but to enable innovation and valorisation (and appropriate the resulting technological options, for example in publications, patents, and as core competence of a start-up firm).

A technology platform is not just a collection of equipment. It enables and constrains further actions. Furthermore, the recognition of the possibility of such platforms incites actions to realize them. As product platform (Gawer and Cusumano, 2002) focuses on the standardisation of interfaces which makes it compatible with the other modules, technological platforms appear as enablers of R&D, of families of technological options, and of successive product development. A sector can then be viewed not in terms of a dominant design and related industry structures, but as a patchwork of technology platforms and related coordination, up to aggregation. Peerbaye (2004) shows how genomics platforms emerged in R&D institutions and some R&D companies (e.g. micro-arrays), but took on a further feature in France when public financing was made available provided there was some geographical concentration and provisions for access ('dispositif instrumental partagé').

In nano R&D and product development, the range runs from the basic set necessary for manipulating at the nanoscale (STM, AFM, surface analysis instrumentation, nano-fabrication including clean room facilities) to further technological (and social) infrastructure necessary for nano-production. This will be different for different types of products: coatings vs. biochips vs. nano-electronics. Such products are not (and most often cannot) be exclusively nano: for example, micro-systems enabled by nano-inputs (components, modifications). When the new industries have become articulated and stabilized, the technology platforms turn into platforms enabling product families in the traditional sense (Tatikonda 1999). What is still distinctive is that these product families are defined by the technology rather than the sector. Start-up companies basing themselves on a technology platform can identify and follow-up opportunities in different sectors.

Technological platforms, when sought after, are intentional opportunity structures. They are also part of evolving (or emerging) techno-industrial networks and help structure them. This note argues that technological agglomeration is the effect of technological platforms being set up, used and expanded. Because of the coordination (*de facto* through the nature of the platform, as well as intentional, e.g. when organizing access) that is involved, there is a proximity effect and some clustering will occur. There are two main routes of technological agglomeration (and one may find other routes in between, a mix of the two main routes).

- building interrelated and interdependent networks, where technological opportunities and platforms get assembled by being available at the same time ("off the shelf"), and allow various exploitations. This can then be recognized for what is happening, optimised, and packaged to be used elsewhere & elsewhen. Already in the region Twente, but definitely the Netherlands (the second case study), one finds a number of nanotechnology value chains (*filières*), some still only emerging. In new fields such a bottom-up fabrication, and to a certain extent bio-nanotechnology, previous arrangements are absent, or are more diffuse. A technological *filière* is not there yet, in contrast to the situation in micro/nano-electronics. Still, one sees technology platforms being constructed and exploited.

- building co-localised facilities and scientific and technological competencies (geographic concentration), where the technology platforms are expansions of existing facilities. They have to be articulated and designed as such, which requires a concerted effort from the beginning. The second route often builds on what has been happening in the first route, in particular when a certain threshold of articulation and stabilization has been passed. The French public policy which supported the creation of technological platforms within the Genopole programme is an example of such articulation allowing further steps to be made (Peerbaye 2004). The Minatec project in Grenoble (our first case study) was conceived as a major new step, but derived its legitimacy from what was happening already in the region.

In both cases, technology platforms need to be located near a research centre or university. The high investment of monetary and human capital into such technology platforms, and the possibility of many various diffuse technology chains to cross at a technological platform, imply that it is attractive to locate the various technology platforms at the same location, near skilled workforce (and a workforce that evolves with the evolution of the technology platform). Small and large companies could then locate themselves nearby and profit from this agglomeration. Platform agglomeration is also an enabling tool to run complementary experiments and to explore different scientific fields. In addition to scientific and technological convergence in nanotechnologies (Roco and Bainbridge, 2002), generic platforms appear to be the locus of hybridization amongst technologies (Avenel *et al.*, 2006), where teams from different traditions and disciplines can meet around technological facilities. Platforms are a hub for the different disciplines to meet (Carlile, 2004), a sharing facility which play the role of a boundary object (Carlile, 2002; Star and Griesemer, 1989).

There will be path dependencies, in the sense that earlier investments and competencies shape what can be done later. Sometimes, such path dependencies are actively constructed by institutional entrepreneurs who mobilize a variety of resources to create a new and major lab (Jean Therme and Minatec in Grenoble) or a distributed set of lab facilities (David Reinhoudt in Twente, and his colleagues in Groningen and Delft, in the Netherlands), which will then have a life of their own. Initiatives from such institutional entrepreneurs will be the other entrance point for our case studies, because these project futures and actively combine resources from different levels. In a particular locality or region, combinations of disciplines and infrastructures can be assembled and exploited that is adapted to existing competencies and networks. For example, Grenoble focuses on nano-electronics and the Twente region in the Netherlands on materials and sensors.

## 3   Illustrative case studies

To explore the agglomeration, we focus on two clusters, Grenoble and the surrounding areas and the cluster/network in the Netherlands. These two cases have been chosen as they are part of the most visible areas involved in nanotechnologies in the world. According to Kahane *et al.*,[2] Grenoble and the Netherlands are two of between 20 and 30 most visible concentrated areas in nanotechnologies, of which nine are in the US and fifteen in Europe (see Figure 1). In the chart (Figure 1), the profiles of the two clusters are quite different as Grenoble exhibits a high specialisation in physics while the Netherlands appears to be rather specialised in biotechnology. For each case study, archival and documentary data were used, including project and funding proposals, consortia agreements, websites, and qualitative and quantitative data on publications and patents. We also interviewed main actors, traced the activities of the promoters of each cluster (Jean Therme and David Reinhoudt), and inventoried firms involved in the clusters and universities.

---

[2] http://www.nanodistrict.org/events/Workshop%20in%20March/nanotec/Kahane.pdf

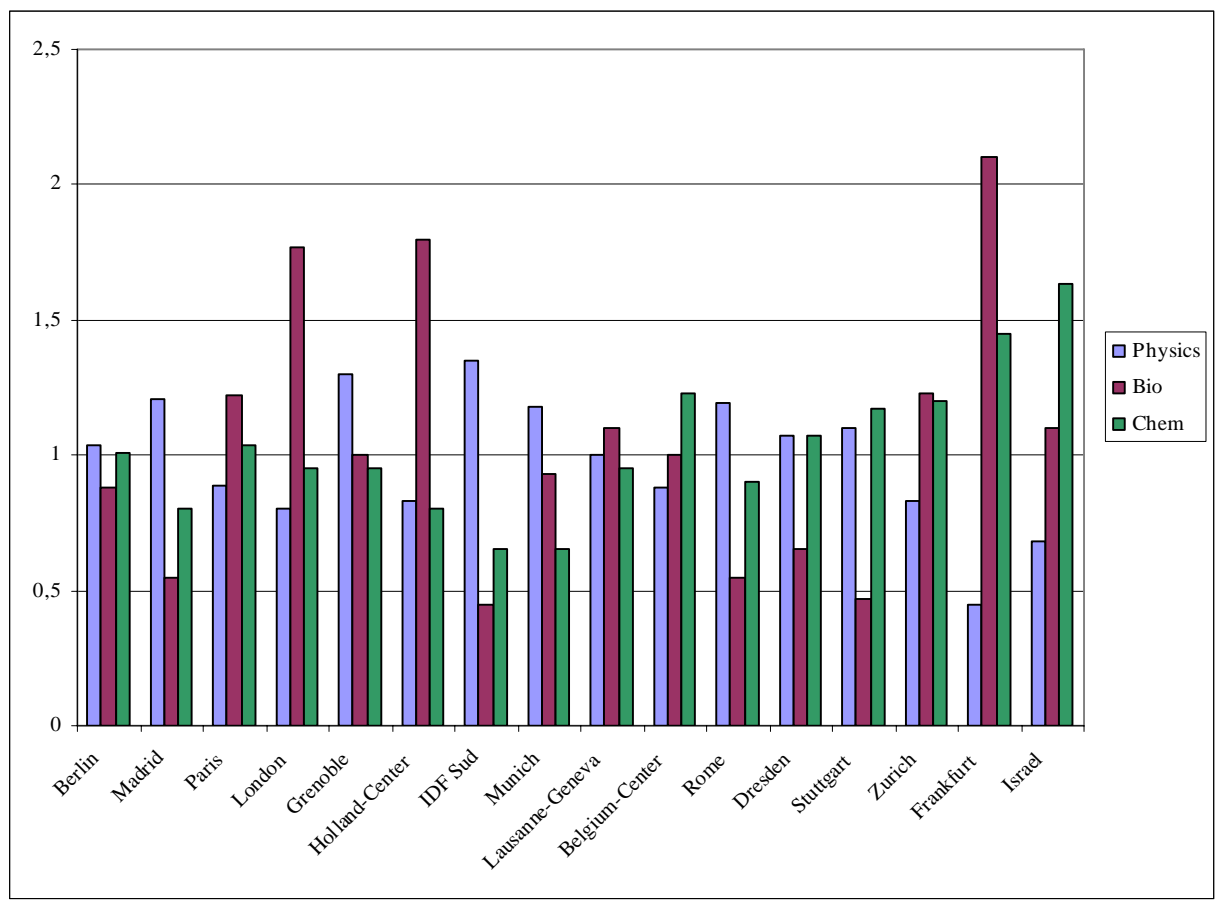

*Figure 1: Main European clusters in nanotechnologies*

## 3.1. Orchestrating technological agglomeration in Grenoble

Technological agglomeration has been occurring in the Grenoble region for a long time. During the early 1980s, LETI (Laboratoire d'Electronique de Technologie de l'Information, a semi-public technological institute dedicated to applied microelectronic research), Thomson Semiconductor (a nationally leading firm at the time) and the Universities of Grenoble formed an alliance to develop research and development capabilities to be able to design and produce wafers of 100mm. They set up shared clean rooms for R&D while production facilities were installed in the neighbourhood of Grenoble to make the transfer of knowledge and know how between R&D and production facilities easier. During the 1990s, the consortium was enlarged to include France

Telecom Research Centre (also located in Grenoble) and to build larger research facilities dedicated to silicon applications, optronics labs and software security (cryptography). In addition, dedicated research and training facilities which belong to different public research organisations (LETI, Universities of Grenoble, European synchrotron research Facility, Leo Langevin Institute) are co-located within the so-called scientific polygon. Micro and nano electronics, structural chemistry, nanobiotechnology, structural biology and generic biotechnology have been developed and formed a local network of interrelated platforms. Actors agreed to share access to the technological platforms and to design rules to manage intellectual property rights, to share the costs of running such platforms (pricing) and to plan the renewal and update of existing facilities as well as the development of new ones. Some of these facilities have been used by start-ups such as Soitec to develop their technologies. So-called 'common labs' between LETI and firms were created later.

In the late 1990s, Minatec was conceived: a new building with shared facilities as well as a collaborative project, promoted by LETI and orchestrated by Jean Therme (Delemarle, 2005), in which the different universities of Grenoble are involved, as well as national labs. Minatec has been formulated as a large and generic scientific and technological facility. It underlines geographic proximity to stimulate scientific and hybridisation amongst the different disciplines which form nanotechnologies. It covers scientific, technological and economic dimensions to support the development of micro- and nanotechnologies. It is not only a hub for scientific teams and firms to collaborate but also an umbrella which groups different the public research organisations. The project was justified by, and could build on four pillars. The first three are a continuum of research organisations, from universities to industry, including LETI as a bridge between basic research and industry; training, with large university campus where engineers and scientists are trained; and a dense network of technology based firms from large multinationals such as Philips or Motorola to recent start-ups like Trixell, Xenocs or Soitec. The fourth pillar is the agglomeration of technological platforms.

The architecture of the building was designed so as to encourage close links between upstream, technology and applied research allocating a central position to technological platforms. Platforms have been assembled in the 20,000 sq meters of Minatec. The actual platforms derive from various groups in the regional scientific (firms and academia) community which opted to share their specific tools of increasing sophistication. Minatec then groups some of them together in the new building, and plans to upgrade them when they are installed in the building during 2006. It also organises platform management facilities and facilitates access to interrelated platforms located in the area. During the resource mobilisation and design phases, centres and their links to other technological resources were already defined, from LETI,[3] and from the region more generally.[4] There is overlapping technological agglomeration. Minatec projected, and now implements, agglomeration of facilities. Characterisation facilities are a further important component, and the idea of "common labs" including special Intellectual Property Right rules was successfully pushed by Jean Therme.

The emergence of Minatec is based on the high concentration of scientific and technological actors. The organisation of the work around the different and coupled technological platforms fosters pluridisciplinarity and problem solving approaches. Minatec emerged from different

---

[3] The Advanced Microelectronics Project Centre (CPMA) enables it to access LETI resources such as the PLATO technology platform (Plasma technology, Lithography: EUV, Nanoimprint, Dielectric materials, Nanomaterials (Si, Ge, Magnetics) and Near field microscopy), the Very Low Temperature Research centre (CRTBT), the Centre for Basic research in condensed materials and the Nanofab which is specialised in the nanofabrication of objects larger than 50nm by particle based (electron and ion beam) lithography, deposition and etching (See Minatec Newsletter, July 2003, at www.minatec.com). It is a keystone of a large number of scientific projects in nano-optics, nanomagnetism or nanoelectronics.

[4] Minatec benefits from the presence of major European facilities, such as Institut Laue Langevin (ILL, neutron source), the European Synchrotron Facility (ESRF), the European Molecular Biology Laboratory (EMBL) and the Grenoble High Magnetic

public research organisations and universities as a hub to produce simultaneously basic research and targeted collaborations with industries. Meanwhile, firms around Grenoble have grown and have decided to realise a joint venture so as to share the costs and the risks in nanoelectronics fabrication. Around SGS Thomson (later to become ST Microelectronics), firms allied to develop a new labfab to produce wafers around 200nm. In 2000, the alliance grew up, including ST Microelectronics, Philips and Motorola to build a new labfab to deal not only with submicronic like in the previous generation but also with nanoelectronics to produce wafers of 200/300 mm. in the same time, one of the world leader in electricity, Schneider Electric decided to set up a new research centre to benefit from the spillovers and from the infrastructure around Grenoble. In 2005, the French government recognized the ensemble which groups Minatec, the fabrication alliance between STMicroelectronics, Philips and Motorola named Crolles 2 and the Schneider new research centre as a world class Pole de competitivité, which implies some preferential treatment.

In the Minatec newsletters (www.minatec.com) there is also reference to the linkages between the research facilities, research and training. About 4,000 employees are to work in Minatec, including 1000 students from Grenoble universities and 2,000 researchers, engineers and teaching staff. Promoted by LETI and universities of Grenoble (especially the engineering, physics and microelectronics departments), it has been positioned as making Grenoble an international centre of nanoscience (Minatec newsletter n°5, January 2004).

This is a success story in the resource mobility and the construction of a rich supply of research and technological opportunity. The question which looms on the horizon is whether to work towards the next integrated set of technological platforms, or to step out of the race altogether.

Field Laboratory (GHMFL) enabling atoms to be observed in fine detail and experiments to be performed which are essential to progress in nanosciences. They are located nearby (less than ½ miles away).

The Crolles 2 production facility is in place; there are some 50 of such facilities worldwide. Actors are already projecting a next "generation", Crolles 3 (of which there will be some 20 worldwide), and negotiate and struggle about what is to be done, and who should take the lead. What we sketched here is the dominant dynamic in the Grenoble region centred around micro- and nano-electronics, one which clearly shows the strong role of technological platforms and evolving industry structures which need nodes where synergies are exploited. There are other activities in the region, e.g. in bionanotechnology. These are much more dispersed but do show signs of emerging technology chains anchored and linked by more or less generic technology platforms. Such a dynamic is clearly visible, and intentionally sought after in our second case, Twente and the Netherlands.

## 3.2 Emerging distributed technological agglomeration in Twente and the Netherlands

Our second case is played out at two levels, regional and national. The geographical scope is perhaps less important for this distinction (the Netherlands is a small country, and could be seen as a region), than the difference in roles of regional actors and authorities, and national level public authorities. The two levels have become linked in two main ways: the mutual positioning of the key nanoscience and technology centres in the Netherlands, and the emergence of a national nanotechnology consortium "NanoNed", which includes a distributed "NanoLab". We shall study the developments in Twente in some detail, as these are centred around a world-level nano-science research institute, MESA+, in the University of Twente, and show some technological agglomeration. For the national consortium, we focus on "NanoLab". There are other interesting aspects including institution building (in which the director of MESA+, David

Reinhoudt, played a major role) and its intended and unintended effects (Mangematin et al. 2005), to which we only refer in passing.

There is mutual positioning of the research institutes, with Groningen as a hub for bionanotechnology, Twente for nanomaterials and manufacture, and Delft for micro- and nano-electronics. We will discuss Twente in more detail below. The Groningen region and University focus on facilities related to preparation, manipulation and detection of cells and biomolecules. In the Technical University of Delft, there is basic nano-science (now organized as a Kavli Institute) as well as work on lithography and nano-electronics, which complements activities of TNO-TPD, a division of the public applied research organisation TNO located in Delft.

Small microtechnology and nanotechnology companies, mainly start-ups, are playing a role in the regions, intertwined with the workings and evolution of the technical platforms. In Twente, where most start-ups are located, they are at the moment both users of facilities and providers of service. Examples include MicronIt, Lionix, and CapilliX, which use the facilities to create micro and nanofluidic platforms for use within the university or by other start-ups, such as Medimate. However, there is still only limited demand for their service in providing tools for R&D. "Killer applications" may arrive, allowing for expansion. None of the bigger firms in the three regions are at present active in nanotechnology, so there is little involvement of what might otherwise be anchor tenants (Agrawal and Cockburn, 2003). There are, of course, non-regional links with big firms like Philips Company.

The history of micro- and nano-research in Twente shows the importance of evolving and overlapping technology platforms. The research institute MESA in the University of Twente, established in 1990, building on an earlier conglomerate of groups and institutes with research in the general area of sensors, actuators and micro-systems. By the end of 1999, further mergers with electronics, optics, and materials research groups led to the establishment of MESA+, with special investments in extensive clean room facilities and linked to a TechPark (itself building on predecessors from the early 1990s). This gradual convergence of fields and the eventual uptake of the 'nanotechnology' banner had much to do with the availability of overlapping technology platforms and the possibility of their expansion – which required institute leaders with particular entrepreneurial characteristics. The competencies built up over the last 20 years include microfabrication, microfluidics and sensors and actuators. MESA+ has high international visibility as it is embedded in networks of excellence, international collaborations and consortia.

For MESA+, spin-offs from the University have become an integral part of micro and nano developments in the region. In the University of Twente research into microfluidics and lab-on-a-chip revolves around the manufacture and manipulation of chip devices both in silica and polymer. Over the last 25 years, University of Twente has built up skills in micromachining to fluidic chips, leading to three spin off companies (LioniX, MicronIt and CapiliX) who develop and produce fluidic chips. The production of the chips occurs in the university cleanroom facilities which are rented by the two companies. Overall, 33% of time of the cleanroom time is rented to companies, limiting the time available for ongoing research at the University of Twente. In addition, 33% of the use of the various technology platforms housed within the MESA+ DANNALAB complex and Central Materials Analysis Laboratory, is allocated to the small companies for characterisation and analysis of products such as pharmaceuticals, nanomaterials, coatings and polymers.

The existence of companies that produce chips on demand, and the mixture of other small companies, which have expertise in thin films, microsieves etc. along with research lines in MESA+ are a further input into an emerging cluster based on (and exploiting) micro and nanofabrication – a national hub and European leader of nanofabrication.

In parallel to these developments, and building on them, a series of initiatives were taken at the national level which would lead, after a number of shifts, to the present R&D consortium NanoNed which draws on government funding. The original aim was to create a stronger position for the three partner centres from the Universities of Twente, Groningen and Delft, in which provision of advanced technical infrastructure was to play a key part. From the 2000 "Masterplan Nanotechnology" onward, a distributed NanoLab, i.e. facilities to be located in the three centres, featured in the plans and proposals. This contains a number of generic technology platforms, not co-located but coordinated across a few locations.

Shifts occurred to address resource mobilisation opportunities, in particular the expansion of the original group of three centres, including, by that time, a division, located in Delft, of the national applied research organization TNO, with centres in four more universities (necessary to avoid accusations of preferential treatment of the original three centres), and eventually also Philips Company. Alignment of the various participants was a challenge, and meeting it (even if precariously) was part of the challenge for the institutional entrepreneurship of David Reinhoudt (Scientific Director of MESA+) in which he was helped by the promise of major funding. Important also was the need to achieve some semblance of coordination between participants who otherwise might see themselves in outright competition. This was done by positioning participants according to their specializations with cross-cutting "flagships" at the consortium national level. NanoLab continued to be a core element, with some 35% of the envisaged resources of the consortium devoted to it. While to be located at the three main centres, it would offer access to other NanoNed participants.

Contrary to Minatec (and Crolles 2) which emphasises co-location to creation a dense cluster of nanotechnologies organised around platforms, the technological agglomeration visible in the so-called NanoLab occurs within dense and highly coordinated networks in the Netherlands. It emphasizes existing competencies and the promise of creating four overlapping generic technology platforms. The Table below shows how the actors themselves described the "hubs" (Figure 2).

| |
|---|
| **Twente: MESA+** focuses on the research and realisation of complex materials, devices and systems, on the processes used for the production of these and on the integration into complex devices and complete systems. Thus it aims to become the Dutch hub for nanofabrication. |
| **Groningen: MSC+ / Biomade** has a fast intensifying focus on the development of (bio)molecular (nano) electronics through a combination of fundamental and applied research. Using the present infrastructure, new functional molecular elements and materials are designed and synthesized. Within the NanoLab NL programme, the MSC+ / Biomade infrastructure is designed to function as: the Dutch centre for bottom-up (bio) molecular electronics and functional (bio) molecular nanostructures. Local organizations putting effort and in such a facility MSC+, Biomade, and the Groningen Academic Hospital (AZG). |
| **Delft: DIMES** has expertise in the field of Micro- and Nano-electronics, mostly using cryogenic techniques, and expertise in Nano-fabrication in many applications.<br><br>With NanoLab NL, DIMES will provide a facility for nano-fabrication for broad use (and for all sorts of material-systems), using high-resolution e-beam lithography, different wet processing, oven-processes, thin film growth, dry-etch, and all sorts of nano-inspection techniques. |
| **Delft: TNO TPD** is primarily focused on production and analysis instrumentation on behalf of mass-fabrication of nano-chips. For this type of research, one needs to be able to measure, develop and experiment on (sub) nanometer scale. Within NanoLab NL the aim is the development of competencies in lithography. |

**Figure 2: Investment and consolidation plan for instrumentation within the NanoLab programme**. (edited version of text from NanoNed proposal to ICES-KIS 2003)

By the end of 2005, NanoLab has invested 20% of its €90 million budget. The project has stimulated larger integration/coordination by the inclusion of Philips NatLab which has now joined NanoLab and is part of the decision making structure for the coordination of investments. Representatives of the five participants (MESA+, DIMES, TNO, MSC+ in Groningen, and Philips) form the board of NanoLab and coordinate the final investments during 2006. This includes the decision for investments, and the fees for use. Thus, it is not just a matter of getting new resources and dividing the spoils. A certain coherence at the level of technical infrastructures is established.

Tensions remain, however, and not just between the university groups. Philips Company, formally part of the NanoNed consortium, continues to pursue its own interests, such as the growth of the research campus it has created on its premises and its avowed goal to push for a micro- and nanotech triangle between Eindhoven (where major research labs are located), Louvain in Belgium (with IMEC) and Aachen in Germany.[5] Since December 2005, the concentration of high tech activities in Eindhoven is recognized by the Dutch government as a "pole de competitivité", and IMEC (Louvain) has established a branch in Eindhoven. The network thickens. And one can speculate about a further form of distributed technological agglomeration, now at the level of the "Low Countries" (Netherlands, Belgium, and the German lower-Rhine region).

## 4 Discussion

While the starting situation and the strategies of key actors are different, the cases of Minatec/Grenoble and Twente/Netherlands both illustrate emerging technological agglomeration. The agglomeration process builds on existing technological competencies, research and training institutions and facilities, but is driven by the recognition of opportunities offered by technological platforms for research as well as for new and existing firms, and by the activities of institutional entrepreneurs mobilising resources for further infrastructure, and creating coordination across actors at the same time. Institutional entrepreneurs like Jean Therme and

---

[5] As Philips Company phrases it: "Initiatives by governments, industries and knowledge institutions are rapidly transforming the region between Aachen, Leuven and Eindhoven from an industry-based area to a technology- and knowledge-based economy with potential to rival some of the world's most prestigious regions of excellence." *Philips Research Password*, 19 (April 2004).

David Reinhoudt have to act at different levels (organizational, regional, national) at the same time. They mobilize support, networks are built and allocation decisions are made, which create a virtual presence of Minatec and NanoLab before actual building occurs. The virtual presence and the promise of new technological opportunities orients actors.

While co-location of the technology platforms is the important and recurrent phenomenon, there are different routes. In Grenoble, in the Minatec project, Jean Therme (and his allies) pools existing infrastructure in the neighbourhood, upgrades those that are needed and adds new ones. In the Netherlands, the strategy of key actors, with David Reinhoudt in the lead, is to reinforce existing competencies by overlaying the facilities with funding for key focal areas, leading to different nano-hubs.

Local arrangements can differ and the 'business models' for the generic platforms must evolve further. In the Netherlands, there are tensions about availability of clean-room time for researchers, dictated by the policy of 33% of the time being made available for small companies. This is compounded by responsibilities of the local hubs to the national NanoLab. In Minatec the organisation of the clean room and related facilities is different: there will be dedicated staff to do fabrication and analysis as a service to a customer. The realisation of actual co-location of equipment from the original institutions and their staff will not be easy though.

The further development may not be conforming to the promises and projections that were made. But it is clear already that there will be effects. Links between universities, public research institutes and firms (small, medium and large) become more important. Regional actors and policy makers become part of the techno-institutional dynamics and changes in industrial networks. Clustering on the basis of technology platforms does not only shape emerging nanotechnology regions, but is also important for the distribution of hubs and Poles de Competitivité at the national level and probably also at the European level. Hybrid roles emerge, for start-ups (see LioniX and MicronIt), and in coordination of facilities with industry (Philips and examples from Minatec) as both users of facilities and providers of a service.

What remains to be clarified is whether this reinforces and balances the creation of clusters based on instrumentation, or whether novel combinations between nano centres, nano networks and nano alliances may appear. The strong claim that agglomeration of technology platforms is a pre-requisite for a nano-cluster needs to be verified further. Further case studies are planned, and while the complexity of developments in the real world will make it difficult to make general claims about factors and drivers, we will disentangle some of the complexity by working with contrasting case studies. The results described above already give an indication that clustering in nanotechnology has interesting dynamics and that the success and failure of a cluster to be stimulated will in part be related to the degree of success in agglomeration of technology platforms.